\documentclass[aps,twocolumn,showpacs,prl]{revtex4-1}
\usepackage{amsmath}
\usepackage{amssymb}
\usepackage{graphicx}
\usepackage{dcolumn}
\usepackage{bm}

\usepackage{bm}
\usepackage{graphicx}
\usepackage{ulem}

\def\be{\begin{equation}}
\def\ee{\end{equation}}
\def\ber{\begin{eqnarray}}
\def\eer{\end{eqnarray}}
\def\bwt{\begin{widetext}}
\def\ewt{\end{widetext}}

\def\nn{\nonumber}

\def\e{{\varepsilon}}
\def\bt{\textbf}

\def\e{\varepsilon}
\def\o{\omega}

\begin{document}

\newcommand{\sgn}{\mathop{\mathrm{sgn}}}
\newcommand{\ef}{\mathop{\varepsilon_{F}}}

\draft

\title {Electron-phonon bound states in graphene in a perpendicular magnetic field}
\author{J. Zhu}
\author{S. M. Badalyan}
\email{Samvel.Badalyan@ua.ac.be}
\author{F. M. Peeters}
\affiliation{Department of Physics, University of Antwerp, Groenenborgerlaan 171, B-2020 Antwerpen, Belgium}

\begin{abstract}
The spectrum of electron-phonon complexes in a monolayer graphene is investigated in the presence of a perpendicular quantizing magnetic field. Despite the small electron-phonon coupling, usual perturbation theory is inapplicable for calculation of the scattering amplitude near the threshold of the optical phonon emission. Our findings beyond perturbation theory show that the true spectrum near the phonon emission threshold is completely governed by new branches, corresponding to bound states of an electron and an optical phonon with a binding  energy of the order of $\alpha\omega_{0}$ where $\alpha$ is the electron-phonon coupling and $\omega_{0}$ the phonon energy. 
\end{abstract}


\maketitle

\paragraph{Introduction --}Observation of the quantum Hall effect \cite{QHE2005Geim,QHE2005Kim} in graphene \cite{Graphene} has stimulated extensive investigations of the quantized motion of massless Dirac fermions in a perpendicular magnetic field \cite{Stormer2007,Carbotte2007,Geim2008,Stroscio2009,Stroscio2010}. Particularly, electron-phonon interaction phenomena have become a subject of active theoretical 
\cite{Ando2007,Falko2007,Falko2009,Ando2011} and experimental \cite{Li2009,Potemski2009,Pinczuk2010,Potemski2011} study. The density of states of both electrons in Landau levels and optical phonons with their weak dispersion shows strong delta-function peaks, which result in sharp resonance structures in the absorption spectra, tunable with the magnetic field and useful for future device applications. 

A unique feature of the graphene single-particle spectrum is that the carrier Fermi velocity $v_{F}\approx 1.15\times10^{6}$ m/s is independent of its energy. As a result, the application of a perpendicular magnetic field $B$ quantizes the electronic Dirac spectrum into nonequidistant chiral Landau levels, $\e_{\mu n}=\mu \sqrt{n} \omega_{B}$, which include also a characteristic zero-energy state \cite{Ando2002,Sharapov2005,Castro2006}. Here $\mu=\pm 1$ denotes the electron chirality and $n=0,1,2,\dots$ the Landau quantum number, $\omega_{B}=\sqrt{2} \hbar v_{F}/ \ell_{B}$ is the magnetic energy and $\ell_{B}$ the magnetic length that determines the localization scale of the electron cyclotron motion. This picture of Landau quantization in graphene is well established, {\it e.g.} by direct observation using scanning tunneling spectroscopy \cite{Li2009,Stroscio2009,Stroscio2010}. 

\begin{figure}[h]
\begin{center}
\begin{minipage}{.225\linewidth}
\centering
\includegraphics[width=\linewidth]{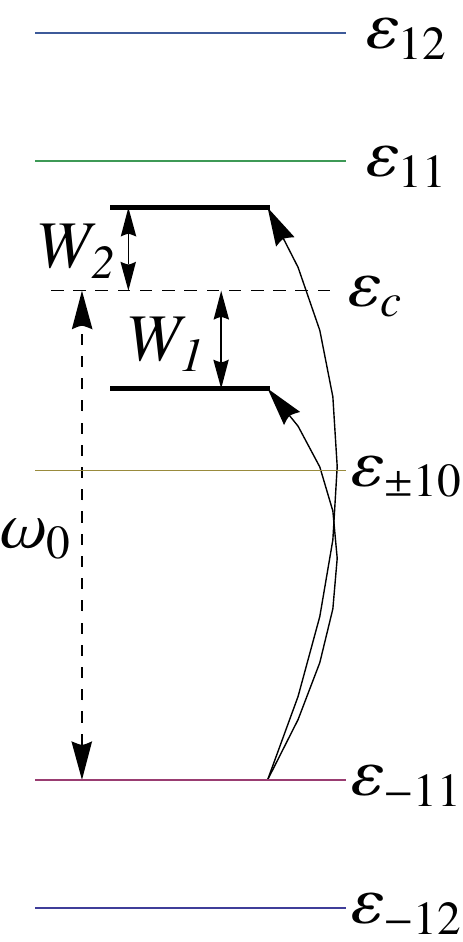}
\end{minipage}%
\hspace{.3cm}
\begin{minipage}{.675\linewidth}
\centering
\includegraphics[width=.95\linewidth]{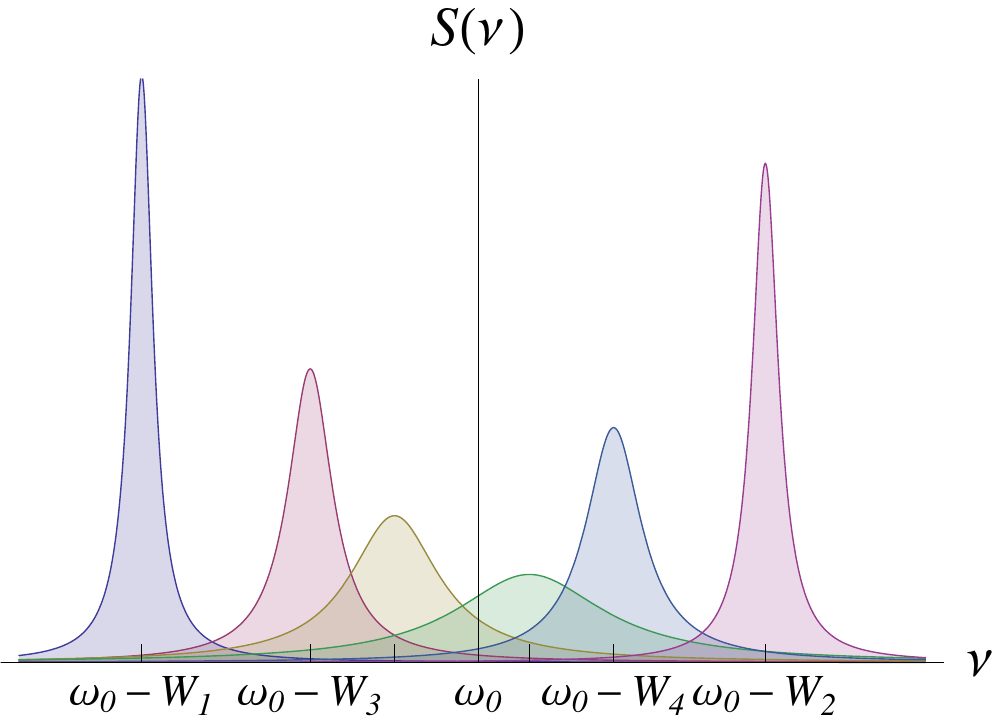}
\par\vspace{0pt}
\end{minipage}
\end{center}
\caption{(left) The electron energy spectrum in graphene, exposed to a perpendicular quantizing magnetic field. The dashed line represents the phonon emission threshold energy $\e^{c}_{\mu n}$. Here it corresponds to the Landau level $\e_{\mu n}$ with $\mu=-1$ and $n=1$. In contrast to the prediction from Ref.~\onlinecite{Nicol2011}, no state corresponds exactly to the energy $\e^{c}_{\mu n}$. Instead, the true spectrum includes a sequence of electron-phonon bound states, which coagulates to the threshold $\e^{c}_{\mu n}$. The most distant bound states with the binding energies $W_{1,2}$ of opposite signs are shown in a larger scale by the bold solid lines below and above $\e^{c}_{-11}$. (right) Schematic diagram of the absorption spectrum in graphene. The cyclotron-phonon resonance is governed by the peaks corresponding to the electron-phonon bound states, which constitute an asymmetric doublet around $\nu=\omega_{0}$.}
\label{fig1}
\end{figure}

In addition to these Landau levels, recently Ref.~\onlinecite{Nicol2011} has proposed a new sequence of discrete states in graphene at energies $\e_{\mu n} \pm \omega_{0}$, corresponding to the emission or absorption of optical phonons with energy $\omega_{0}$. According to this concept, based on perturbation theory, phonon-assisted electron transitions between two different Landau levels can manifest themselves as peaks in the density of states at frequencies $\e_{\mu' n'}-\e_{\mu n}\pm \omega_{0}$. These transitions involve optical phonons, which exist but do not interact with electrons. In the present paper we show that even for small electron-phonon coupling, $\alpha$, the true spectrum in graphene in a perpendicular quantizing magnetic field develops a fine structure in the neighborhood of the phonon emission threshold, $\e^{c}_{\mu n}=\mu \sqrt{n} \omega_{B} + \omega_{0}$, due to the formation of new spectral branches, corresponding to {\it bound states} of an electron and of an optical phonon. Therefore, electronic transitions should involve these bound states and should occur not at the phonon emission threshold energy $\e^{c}_{\mu n}$ but above and below it at the characteristic scale of the binding energy, $W_{gr}=\alpha\omega_{0}$ (see Fig.~\ref{fig1}(left)) \cite{foot1}. The bound states of electron and optical phonon were first proposed for bulk semiconductors in Ref.~\onlinecite{Levinson1970} and it is important to note that no resonance situation is required for the formation of the bound states. In graphene the electron-phonon binding energies are of both signs because no continuum exists in the electron+phonon system in the presence of the magnetic field and the bound states appear both below and above the phonon emission ``threshold''.   
The strength of electron-phonon coupling in graphene $\alpha\sim 0.02$ \cite{tse2} while the phonon energy $\omega_{0}\approx 196$ meV and we find that the binding energy should be of the order of $W_{gr}=4$ meV. Because of its different functional dependence on the coupling constant $\alpha$, the the binding energy in graphene is much larger than that in bulk samples \cite{1Kaplan1972} or in graphene samples in the absence of a magnetic field \cite{2SMB2012}.
In conventional two dimensional electron systems the binding energy scale in a magnetic field has the same functional dependence on $\alpha$ \cite{1SMB1988}, however due to the larger phonon energy, $W_{gr}$ is substantially larger in graphene than binding energies in InSb ($W_{2D}=0.5$ meV) and GaAs ($W_{2D}=2.5$ meV) structures, extensively studied in experiment \cite{Littler1989,Bruno2001}. Our calculations show that the dependence of the binding energy in graphene on the magnetic field is weaker than that in conventional systems and for $\o_{B}\sim\o_{0}$ actual binding energies vary within $5-15$ meV that is fully measurable. Moreover in graphene in such moderate fields, the bound states can be located between the zero and first Landau levels with $\mu=+1$ (see Fig.~\ref{fig1}(left)) and this creates an additional possibility that the splitting of the absorption peak can also be observed in combined resonance absorption with a participation of the zero Landau level. In this case the light frequency $\o$ can substantially differ from the frequency of transverse optical phonons, near which the lattice reflection is important \cite{Summers1968,Levinson1973}. This in its turn facilitates the experimental observation of the bound states in graphene.

In experiment the bound states will manifest themselves in the absorption spectrum of the cyclotron-phonon resonance \cite{Bass1965,McCombe1967}. Upon absorbing a photon of frequency close to $\nu=\e_{\mu' n'}-\e_{\mu n} + \omega_{0}$, an electron is transferred from the initial Landau level $\mu n$ to the final state $\mu' n'$ and simultaneously it creates an optical phonon, which is bound to the electron in its final state. This process results in a {\it fine structure} of the cyclotron-phonon resonance spectrum $S(\nu)$, which for $n=n'$ and $\mu=\mu'$ is shown in Fig.~\ref{fig1}(right). Instead of a single delta-function peak at $\nu=\omega_{0}$, the absorption spectrum constitutes an asymmetric doublet around the phonon frequency $\omega_{0}$ with a splitting of the peak determined by $\nu-\omega_{0}\sim W_{gr}$ \cite{foot2}.

\begin{figure}[t]
\begin{center}
\includegraphics[width=.6\linewidth]{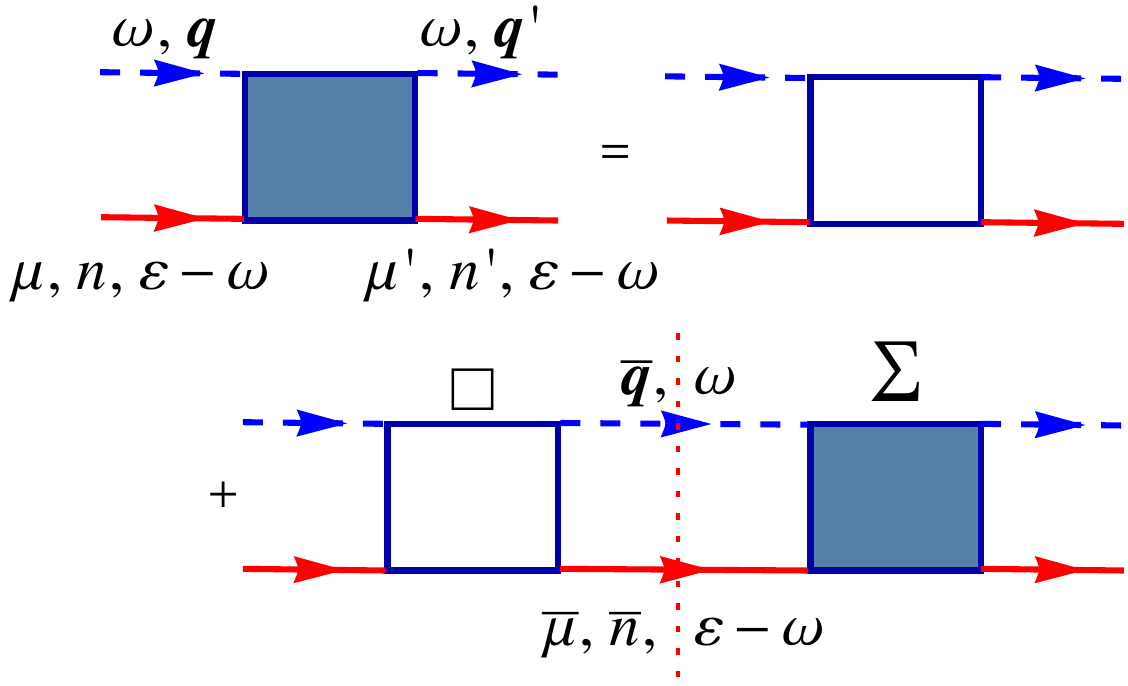} \\ \par\vspace{5pt}
\includegraphics[width=.9\linewidth]{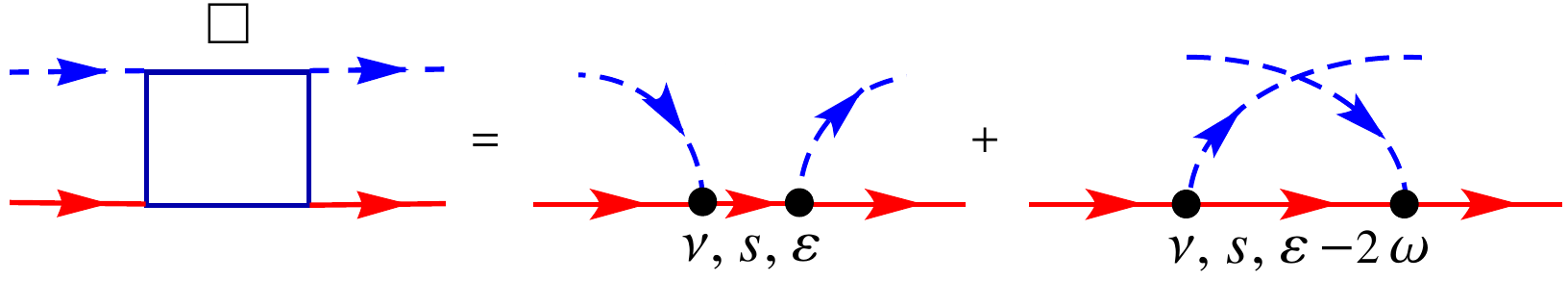}
\end{center}
\caption{ Equations for the electron-phonon scattering amplitude. (top) The exact scattering amplitude $\Sigma(\e)$ with a "dangerous" intersection with respect to one electron and one phonon line. (bottom) The irreducible four vertex part $\square(\e)$, approximated by the bare amplitude. The solid and dashed lines correspond to the electron and phonon Green functions, the bold dots to the bare electron-phonon vertices. The vertical dotted line shows the "dangerous" intersection.}
\label{fig2}
\end{figure}

\paragraph{Threshold approximation and diagrammatic technique --}We calculate the spectrum of the electron-phonon bound states from the poles of the electron-phonon scattering amplitude $\Sigma(\e)$ with respect to the energy parameter $\e$, corresponding to the total energy of the electron and phonon \cite{foot3}. Despite the weak bare electron-phonon coupling, the effective electron-phonon interaction is strong in the energy range close to the phonon emission threshold $\e^{c}_{\mu n}$ where the density of final states of the electron+phonon system diverges. Therefore, the calculation of $\Sigma(\e)$ by means of perturbation theory is not applicable [see the review papers in Ref.~\onlinecite{Levinson1973}]. The exact electron-phonon scattering amplitude satisfies the Dyson type integral equation, drawn in Fig.~\ref{fig2}, in which we have explicitly separated a "dangerous" intersection near the threshold with respect to one electron and one phonon line. Such an intersection corresponds to the emission of an almost real optical phonon and results in a singular term, which is responsible for the enhancement of the effective electron-phonon coupling and for the formation of the electron-phonon bound states.

The diagrammatic equations in Fig.~\ref{fig2} are depicted in a gauge invariant diagrammatic technique (details about this technique can be found in the Supplementary information and in Refs.~\onlinecite{Levinson1971,Levinson1973}). The filled and open squares correspond to the exact, $\Sigma_{\mu n,\mu' n'}(\e | {\bf q}, {\bf q^\prime})$, and the bare, $\square_{\mu n,\mu' n'}(\e | {\bf q}, {\bf q^\prime})$, electron-phonon scattering amplitudes, the internal solid and dashed lines to the exact electron and phonon Green functions, $ G_{\mu n}(\e)$ and $D(\o,{\bf q})$. 
All these quantities do not depend on the electron gauge non-invariant quantum number $k_{y}$, {\it i.e.} on the choice of the vector potential gauge. 
In the "dangerous" intersection in the second term of the rhs of the diagrammatic equation for $\Sigma(\e)$ in Fig.~\ref{fig2}, we can at low temperatures replace the exact phonon Green function by a free phonon propagator. After taking the integration over the phonon energetic parameter $\omega$, it is replaced by the phonon energy $\omega_{0}$. 
Then, the integral equation for $\Sigma(\e)$ in Fig.~\ref{fig2} can be written analytically as 
\begin{eqnarray}\label{iesa}
\Sigma_{\mu n,\mu' n'}(\e | {\bf q}, {\bf q^\prime}) = \square_{\mu n,\mu' n'}({\bf q},{\bf q'})
+\sum_{\bar{\mu}, \bar{n}, \bar{\bf q}} \square_{\mu n,\bar{\mu} \bar{n}}({\bf q},\bar{\bf q})
\\
\times
G_{\bar{\mu}\bar{n}}(\e-\omega_{0})\exp\left(-i \Delta\chi \right) \Sigma_{\bar{\mu} \bar{n},\mu' n'}(\e | \bar{\bf q}, {\bf q'})~.\nn
\end{eqnarray}
Here the exponential factor appears due to the gauge non-invariant part of the electron-phonon vertex (see Eq.~\ref{ephvp} and the Supplementary information for details). The phase $\Delta\chi=\ell^{2}_{B}[{\bf q}-\bar{\bf q},\bar{\bf q}-{\bf q'}]/2$ where the square brackets denote the out of plane component of the vector product.

In the energy range of our interest $\e \approx \e^{c}_{\mu n}$ the energy parameter $\e-\omega_{0}$ of the electron Green function in the second term of the rhs of Eq.~(\ref{iesa}) is far from the threshold $\e^{c}_{\mu n}$, therefore the exact electron Green function can be taken without allowance for interaction with phonons and be replaced by the bare function $G^{0}_{\bar{\mu}\bar{n}}(\e-\omega_{0})$. In the Landau representation $G^{0}_{\mu n}(\e)$ is diagonal 
and is given by its usual pole form 
\begin{eqnarray}
G^{0}_{\mu n}(\e)=\left(\e- \e_{\mu n}+i 0 \cdot \sgn{\e}\right)^{-1}~.
\label{egf}
\end{eqnarray}
Therefore, in the sum in Eq.~(\ref{iesa}) over the electron internal quantum numbers $\bar{\mu}, \bar{n}$ only the singular term with $\bar{\mu}=\mu$ and $\bar{n}=n$ should be retained in the neighborhood of the phonon emission threshold $\e \approx \e^{c}_{\mu n}$.

The central quantity in Eq.~(\ref{iesa}) is the bare amplitude $\square_{\mu n,\bar{\mu}\bar{n}}({\bf q},{\bf \bar{q}})$, which has no "dangerous" intersection with respect to one phonon and one electron line and therefore can be calculated within perturbation theory. In the lowest order in $\alpha $, the two diagrams shown in Fig.~\ref{fig2} contribute to $\square _{\mu n,\bar{\mu}\bar{n}}({\bf q},{\bf \bar{q}})$. The bold dots represent the bare electron-phonon vertices $\gamma({\bf q})$, which we calculate in Supplementary information. For longitudinal optical phonons we find
%
%
%
%
\begin{eqnarray}\label{ephvp}
\gamma _{\mu nk_{y};\mu ^{\prime }n^{\prime }k_{y}^{\prime }}(\mathbf{q}) &=&\sqrt{\alpha }\frac{\hbar v_{F}}{\mathcal{L}}\delta _{k_{y}^{\prime},k_{y}+q_{x}}\exp \left[ i\phi _{\mathbf{q}}(n-n^{\prime })\right]  \notag
 \\
&\times &\exp \left[ iq_{x}(k_{y}+k_{y}^{\prime })/2\right] P_{\mu n,\mu^{\prime }n^{\prime }}(t)
\end{eqnarray}%
where $\mathcal{L}$ is the normalization length, $\phi=\phi_{\bf q}$ the polar angles of the vector ${\bf q}$. The gauge-invariant part of the vertex is 
\begin{equation}
P_{\mu n,\mu ^{\prime }n^{\prime }}(t)=(-1)^{n-n^{\prime}}c_{n}c_{n^{\prime }}\left[\mu ^{\prime}Q_{n,n^{\prime }-1}(t)-\mu Q_{n-1,n^{\prime }}(t) \right]~.
\end{equation}%
with $c_{0}=1$ and $c_{n}=1/\sqrt{2}$ for $n\geq1 $. For $n\geq m$ the form factor $Q_{nm}(t)=\sqrt{m!/n!}e^{-t/2}t^{(n-m)/2}L_{m}^{n-m}(t)$ where $L_{m}^{n}(t)$ is the Laguerre polynomial and $t=q^{2}\ell _{B}^{2}/2$. For $n<m$, one can use the identity $Q_{nm}(t)=(-1)^{n-m}Q_{mn}(t)$. For either $n=0$ or $m=0$ the form factor $Q_{nm}(t)$ with a negative index is identically zero, which means that in graphene the electron-phonon matrix elements between the zero-energy Landau states vanish identically, $P_{\mu 0,\mu ^{\prime }0}(t)\equiv 0$.

As seen from Eq.~(\ref{ephvp}), the gauge non-invariant part of the electron-phonon interaction vertex in graphene is the same as the one in conventional electron systems with parabolic dispersion \cite{LevGant}. Consequently, it allows us to apply the rules of the gauge invariant diagrammatic technique for conventional systems from Ref.~\onlinecite{Levinson1971} when calculating the electron-phonon scattering amplitude $\Sigma(\e)$ in graphene. We just replace the gauge invariant form factor $Q_{n m}(t)$ by $P_{\mu n, \mu' n'}(t)$.

\paragraph{Dispersion equation of electron-phonon bound states --}To find the spectrum of the bound states it suffices to consider the amplitude $\Sigma(\e)$ with the $\mu=\mu'$ and $n=n'$ external electron lines. After the above simplifications we can also eliminate the phase factor $\Delta\chi$ in Eq.~(\ref{iesa}) by introducing a symmetric definition for the scattering amplitude $\Sigma'(\e | {\bf q, q'})=\exp\left(-i\frac{\ell^{2}_{B}}{2}[{\bf q},{\bf q'}]\right) \Sigma(\e | {\bf q, q'})$ (similarly for $\square'({\bf q, q'})$). These primed quantities will depend only on the difference of the polar angles $\phi=\phi_{\bf q}-\phi_{\bf q'}$ of the vectors ${\bf q}$ and ${\bf q'}$. Therefore, we find that the Fourier components $\Sigma_{\mu n}'^{l}(\e; t t')=\int_{0}^{2\pi} d\phi/(2\pi) \exp(-i l \phi) \Sigma_{\mu n}'(\e | {\bf q, q'})$ of the new amplitudes with different $l=0,\pm1,\pm2,\dots$ satisfy the independent integral equations 
\begin{eqnarray}\label{ie}
R_{\mu n}^{l}(\e | t, t') &=& K_{\mu n}^{l}(t,t')\\
&+&\Lambda_{\mu n}(\e)\int_{0}^{\infty}d\bar{t}K_{\mu n}^{l}(t,\bar{t})R_{\mu n}^{l}(\e | \bar{t},t')~. \nn
\end{eqnarray}
Here we define the dimensionless amplitudes, $R_{\mu n}^{l}(\e | t, t')=(\pi \gamma/\tilde{\alpha}\omega_{0})\Sigma'^{l}_{\mu n}(\e | t, t')$ and $K_{\mu n}^{l}(\e | t, t')=(\pi \gamma/\tilde{\alpha}\omega_{0}) \square'^{l}_{\mu n}(t, t')$ where $\gamma=g_{s}g_{v}{\cal L}^{2}/2\pi\ell^{2}_{b}$ is the capacitance of the Landau level allowing for two spin orientations, $g_{s}=2$, and for the valley degeneracy, $g_{v}=2$. We introduce also the function 
\begin{eqnarray}\label{Lambda}
\Lambda_{\mu n}(\e)=\frac{\tilde{\alpha}}{4\pi} \frac{\omega_{0}}{\e-\e^{c}_{\mu n }}~,
\end{eqnarray}
which determines the effective electron-phonon interaction in the energy region near the threshold $\e^{c}_{\mu n }$ with the renormalized coupling strength $\tilde{\alpha}=\alpha\omega_{B}/\omega_{0}$. As seen the amplitude $R_{\mu n}^{l}(\e | t, t')$ is the resolvent of the inhomogenous Fredholm integral equation (\ref{ie}) with the kernel $K_{\mu n}^{l}(t,t')$, which we can write explicitly as
\bwt
\ber\label{kern}
K_{\mu n}^{l}(t,t')=\sum^{\infty}_{m=0, \nu=\pm1} (-1)^{s-n}P_{\mu n,\nu m}(t)P_{\nu m, \mu n}(t')\left[\frac{J_{l+m-n}(2\sqrt{t t'})}{(\nu\sqrt{m}-\mu\sqrt{n})+\sigma}+\frac{\delta_{l,m-n}}{(\nu\sqrt{m}-\mu\sqrt{n})-\sigma} \right]~.
\eer
\ewt
Here $\sigma=\omega_{0}/\omega_{B}$, $J_{l}(t)$ is the Bessel function, and $\delta_{l,m}$ is the Kronecker delta. Hence the resolvent $R_{\mu n}^{l}(\e | t, t')$ has a pole in $\e$ when $\Lambda_{\mu n }(\e)=\lambda^{l}_{\mu n r}$ where $\lambda^{l}_{\mu n r}$ are eigenvalues of the kernel (\ref{kern}), numbered by the index $r=1,2,\dots$ for a given $l$. Thus, the energies of the electron-phonon bound states are given by
\begin{eqnarray}
{\e}_{\mu n r}^{l}={\e}^{c}_{\mu n }-\tilde{\alpha} \frac{\omega_{0}}{\lambda^{l}_{\mu n r}} \equiv {\e}^{c}_{\mu n } - W^{l}_{\mu n r}
\label{de}
\end{eqnarray}
where $W_{\mu n r}^{l}$ defines the binding energy of the bound states, which refer to the threshold $\e^{c}_{\mu n}$. As far as there is no continuum in the bare spectrum, we find that the new branches of the spectrum are not damped and appear in the spectrum both below ($W^{l}_{\mu n r}>0$) and above ($W^{l}_{\mu n r}<0$) the ``threshold`` energy $\e^{c}_{\mu n }$, {\it i.e.} the binding energies are determined by the kernel eigenvalues $\lambda^{l}_{\mu n r}$ of both signs.
\begin{table}[h]
\vspace{-0.3cm}
\caption{
Eigenvalues of $\kappa _{r}^{l}=\left( -1\right) ^{l}\kappa _{r}^{-l}$ in strong magnetic fields.}
\label{tab1}
\begin{ruledtabular}
\begin{tabular}{ccccccc}
$\kappa _{r}^{l}$ &  $r=1$ & $2$ & $3$ & $4$ & $5$
& $6$ \\ 
$l=0$ & $0.97$ & $-5.08$ & $9.96$ & $-23.49$ & $53.68$ & $-126.5$\\ 
$1$ & $-3.27$ & $5.19$ & $-10.24$ & $21.22$ & $-46.19$ & $103.4$ \\ 
$2$ & $3.42$ & $-6.38$ &$13.23$ & $-28.68$ & $64.14$ & $-146.6$\\
\end{tabular}
\end{ruledtabular}
\end{table}
The bound states are identified by the total angular momentum $l$ of the electron rotation around the magnetic field, which takes both non-negative and negative integer values. For a given $l$ there exists a sequence of bound states, labeled by the index $r$. The latter emerges due to the integral nature of the equation for the scattering amplitude $\Sigma(\e)$. Because the dispersion of the optical phonon is negligibly small, the energy in the "dangerous" intersection does not depend on the phonon momentum $q$: all intersections with different $q$ are singular and phonons with any $q$ participate in the formation of the bound states.

\paragraph{Binding energies of electron-phonon bound states --}
The analysis of the eigenvalue problem for the kernel (\ref{kern}) shows that in high magnetic fields ($\sigma\ll 1$) the leading order in $\sigma^{-1}$ contribution to the sum in (\ref{kern}) comes from the term $\nu=\mu$ and $m=n$. Hence, in this regime the effective electron-phonon coupling for electrons from the $n=1$ Landau level, $P_{\mu 1,\mu 1}(t)$, (recall for $n=0$ the electron-phonon vertex $P_{\mu 0,\mu 0}(t)=0$) determines the kernel
\ber\label{kern3}
K_{\mu n}^{l}(t,t')=\frac{1}{\sigma}Q_{10}(t)Q_{10}(t')\left[J_{l}(2\sqrt{t t'})-\delta_{l,0}\right]~,
\eer
which is independent of the chirality $\mu$. The eigenvalues even of this simplified kernel cannot be found analytically. However, the parameter $\sigma$ enters in the kernel as a factor so its eigenvalues are given by $\lambda^{l}_{\pm 1r}=\sigma \kappa^{l}_{r}$ where $\kappa^{l}_{r}$ are the eigenvalues of the kernel $\sqrt{t t'}\exp{\left[-(t+t')/2\right]}\left[J_{l}(2\sqrt{t t'})-\delta_{l,0}\right]$, which we calculate numerically and summarize in Table~\ref{tab1}. Thus, in this strong field regime the binding energies are given by $W^{l}_{r}=\alpha\omega_{0}/(\sigma^{2} \kappa_{r}^{l})$, which increase linearly with $B$ and decrease rapidly with $r$, asymmetrically on both sides of the threshold. 
In weak magnetic fields ($\sigma\gg 1$) the spacing $(\sqrt{m+1}-\sqrt{m})\omega_{B}\sim \omega_{0}/2\sigma^{2}$ between adjacent Landau levels $m$ and $m+1$ around the phonon emission threshold becomes smaller than the scale of the binding energy $W\sim \alpha\omega_{0}$. Therefore, in graphene already for $1/\sqrt{2\alpha}\lesssim \sigma \sim \sqrt{m} $ Landau levels form a quasi-continuum, which hinders the formation of the bound states in such weak fields. 

\begin{figure}[t]
\begin{center}
\includegraphics[width=0.75\linewidth]{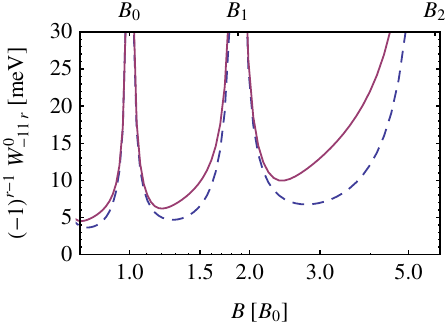}
\end{center}
\caption{The binding energies $W_{\mu n r}^{0}$ for the bound states, which refer to the threshold $\e^{c}_{\mu n}$  with $n=1$ and $\mu=-1$. The solid and dashed curves correspond to the $r=1$ and $2$ bound states, which are most distant from the threshold.} 
\label{fig3}
\end{figure}

For arbitrary magnetic fields we calculate the binding energy of bound states numerically from the full kernel (\ref{kern}). Near the resonances  $\e_{\mu' n'}-\e_{\mu n}=\pm \omega_{0}$ between the bare graphene Landau levels, electron-phonon hybrid states are formed in the spectrum, which have a much larger binding energy scale of the order of $\sqrt{\alpha}\omega_{0}$. These resonance states can be obtained within perturbation theory from the poles of the single- or two-particle Green function (but from the poles of the scattering amplitude) and determine the structure of the magnetophonon resonance. This effect has been recently studied in Refs.~\onlinecite{Ando2007,Falko2007} and is here excluded from the consideration. Far from the resonances the spectrum is completely governed by the electron-phonon bound states (\ref{de}), which determine the fine structure of the cyclotron-phonon resonance ({\it cf.} Fig.~\ref{fig1} (right)). In Fig.~\ref{fig3} we plot the binding energies $W^{0}_{-1 1 r}$ as a function of the magnetic field strength $B$ for the most distant $r=1$ and $2$ electron-phonon bound states with $l=0$, referring to the threshold $\e^{c}_{-1 1}$. The divergences of the binding energy correspond to the hybrid states at the resonance fields $B=B_{0}$, $B=B_{1}$, and $B=B_{2}$. Here $B_{0}\approx 22.04$ T, $B_{1}\approx 1.87 B_{0}$ and $B_{2}\approx 5.83 B_{0}$ are given, respectively, by the resonance conditions  $\omega_{0}=\omega_{B}$, $\omega_{0}=(\sqrt{3}-1)\omega_{B}$, and $\omega_{0}=(\sqrt{2}-1)\omega_{B}$. It is seen that for intermediate magnetic fields far from the resonances $B_{0}$, $B_{1}$, and $B_{2}$, the actual binding energies of the electron-phonon bound states vary approximate from $5$ to $15$ meV and are measurable in light absorption experiments.

In conclusion, we have calculated the electronic spectrum in a monolayer graphene in a perpendicular magnetic field. It is shown that the true spectrum near the optical phonon emission threshold consists of new branches of electron-phonon bound states, which will manifest themselves in the absorption spectrum as an asymmetric doublet around the threshold energy.  

We acknowledge support from the Belgian Science Policy (BELSPO and EU), the ESF EuroGRAPHENE project CONGRAN, and the Flemisch Science Foundation (FWO-Vl).


\begin{thebibliography}{99}
\bibitem{foot1} Here we assume that graphene is at low temperature so that we can neglect phonon absorption processes.
\bibitem{foot2} Theory of the fine structure of the cyclotron-phonon resonance in graphene will be reported elsewhere.
\bibitem{foot3} Because the number of phonons is not conserved, the spectrum obtained from the poles of $\Sigma(\e)$ and of the single-particle Green function are the same.
\end{thebibliography}
\end{document}